\documentclass{article}
\usepackage{spconf,amsmath,graphicx, mathtools}
\usepackage{amssymb}
\usepackage{booktabs}
\usepackage{subfig}
\usepackage[english]{babel}
\usepackage{xcolor}
\usepackage{makecell}
\usepackage{varwidth}
\usepackage{xcolor}
\usepackage{paralist}
\usepackage{bbm}
\usepackage{multirow}
\usepackage{hyperref}
\usepackage{breqn}
\usepackage{longtable}
\usepackage{mathtools}
\usepackage{algorithm}
\usepackage{algpseudocode}
\usepackage{paralist}
\usepackage{microtype}
\usepackage[detect-all]{siunitx}
\usepackage{cleveref}
\usepackage{lscape} 
\usepackage{upgreek}

\title{Active Learning of Non-semantic Speech Tasks with Pretrained Models}
\name{Harlin Lee$^1$\thanks{This work was partially supported by NSF grants DMS-1952339 and
DMS-2152717.  It was also partially supported by NGA award
HM0476-21-1-0003, approved for public release, NGA-U-2022-02425; any
opinions, findings and conclusions or recommendations expressed in this material are those of the authors and do not necessarily reflect the views of the NGA.  Email: harlin@math.ucla.edu, a.saeed@tue.nl, bertozzi@math.ucla.edu. Icons are from TheNounProject.}, Aaqib Saeed$^2$, Andrea L. Bertozzi$^1$}
\address{$^1$University of California Los Angeles, Los Angeles, CA, USA \\
$^2$Eindhoven University of Technology, Eindhoven, The Netherlands
}

\begin{document}
\maketitle

\begin{abstract}
Pretraining neural networks with massive unlabeled datasets has become popular as it equips the deep models with a better prior to solve downstream tasks. However, this approach generally assumes that the downstream tasks have access to annotated data of sufficient size. In this work, we propose ALOE, a novel system for improving the data- and label-efficiency of non-semantic speech tasks with active learning (AL). ALOE uses pretrained models in conjunction with active learning to label data incrementally and learn classifiers for downstream tasks, thereby mitigating the need to acquire labeled data beforehand. We demonstrate the effectiveness of ALOE on a wide range of tasks, uncertainty-based acquisition functions, and model architectures. Training a linear classifier on top of a frozen encoder with ALOE is shown to achieve performance similar to several baselines that utilize the entire labeled data.
\end{abstract}


\section{Introduction}
\label{sec:introduction}

Deep neural networks require a large amount of well-annotated training data to generalize well. In the real world, access to labeled datasets is limited, and collecting abundant examples requires significant investment in terms of both finance and time. Further, the expertise required to collect high-quality labels can be limited in domains like health monitoring. Pretraining neural networks with massive unlabeled datasets have become a popular choice to tackle this issue, as it equips the deep models with a better prior for downstream problems. In particular, self-supervision has been shown to achieve tremendous success across data modalities including audio and speech \cite{shor20interspeech, baevski2020wav2vec, saeed2021contrastive}. Self-supervised learning tasks a neural network to solve an auxiliary learning problem for which supervision can be acquired from the unlabeled input itself. This pushes the model to learn useful representations from unlabeled data. One can then use the pretrained model (or an encoder) either as a fixed feature extractor or as initialization in transfer learning. %

Still, the downstream learning tasks require sufficient annotated samples for one to train a classifier on top of the encoder from the self-supervised or unsupervised pretraining phase. %
We note that for some tasks, pretrained models may require only a small percentage (e.g., $10$\%-$20$\%) of labeled data to match the performance of a supervised model, but the number of labeled instances can still be huge. As mentioned previously, data labeling is cumbersome and expensive, and it is unclear a priori for which instances we should get annotations, as not all examples carry useful information for learning. In comparison, getting unlabeled data for the desired task is much easier. 

\begin{figure}[t]
\centering
\includegraphics[width=0.95\columnwidth]{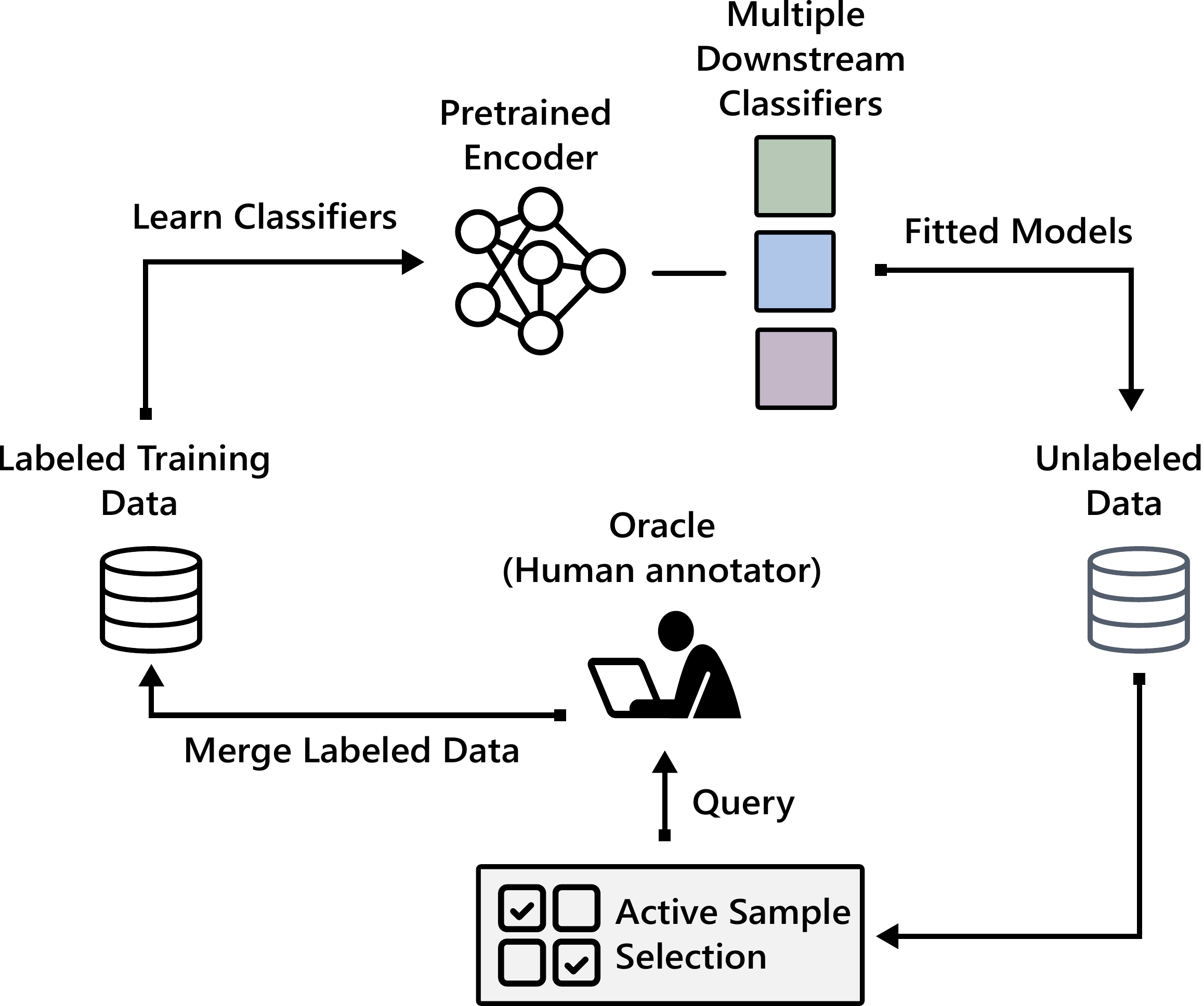} 
\caption{\small{Illustration of ALOE: active learning with a pretrained model. The encoder parameters are frozen, allowing the use of the same encoder across multiple downstream tasks.}}
\label{fig:aloe_overview}
\end{figure}

In light of these observations and challenges, we design ALOE\footnote{Code: \url{https://github.com/HarlinLee/ALOE}} (\textbf{A}ctive \textbf{L}earning of n\textbf{o}n-s\textbf{e}mantic speech tasks with pretrained models) to improve the label- and data-efficiency of non-semantic speech tasks~\cite{shor20interspeech}. Figure \ref{fig:aloe_overview} describes how ALOE exploits pretrained models with \textit{active learning} (AL) for gradual labeling of data and learning classifiers for downstream tasks. While AL has been explored in speech recognition \cite{Riccardi2005,jiaji2016active,malhotra19_interspeech}, to the best of our knowledge, ALOE is a first attempt at improving label efficiency for non-semantic speech tasks with AL and pretrained models. In principle, AL provides a systematic way to procure labels for the most informative instances, e.g., a class for which a model is highly uncertain \cite{Lewis94asequential,settles2012active}. AL framework comprises three stages: a) acquisition---selecting relevant examples, b) annotation---getting ground-truth labels from a human annotator; and c) retraining---learning model on acquired and existing labeled instances. In practice, these steps are time-consuming and computationally expensive because a model has to be trained many times during the AL cycle. We improve the efficiency of AL with the power of semantic representations from pretrained models while also enhancing its acquisition capability, which is otherwise limited when training a model from scratch. Furthermore, we are privy to a better view of the annotation process with human-in-the-loop.     

We demonstrate the effectiveness of our approach on a broad range of tasks, uncertainty-based acquisition functions, and model architectures. Training a linear classifier on top of a frozen encoder with AL uses only a fraction of the examples yet results in a similar performance as several baselines that utilize the entire labeled data. ALOE provides an end-to-end system for exploiting pretrained models more effectively and mitigates the need to acquire large labeled data beforehand. 

\section{Methodology}
\label{sec:methodology}

We propose to use a pretrained (unsupervised or self-supervised) model with active learning (AL) to improve the data- and label-efficiency of deep models in non-semantic speech tasks. Importantly, ALOE has a shared fixed pretrained encoder with separate shallow classifiers for each end-task. Our approach leverages generic speech representations learned from massive amounts of unlabeled data and identifies key samples from an intended end-task that are to be labeled by a human annotator (i.e., an oracle) depending on their uncertainty scores. 

We consider an AL setup with the pool-based acquisition as commonly studied in the literature~\cite{settles2012active,miller2022graph}. Let $ \mathcal{X} \subseteq \mathbb{R}^d$ be $d$-dimensional speech data samples, $\mathcal{Y}$ be labels of a non-semantic speech classification task, and $\mathcal{F}_{\theta}: \mathcal{X} \to \mathbb{R}^m$ be a pretrained encoder, whose embedding dimension $m \ll d$. Given some labeled instances $\mathcal{D}_{l} \subset \mathcal{X} \times \mathcal{Y}$ and a large amount of unlabeled data $\mathcal{D}_{u} \subset \mathcal{X}$ with $|\mathcal{D}_{l}| \ll |\mathcal{D}_{u}|$, the aim is to incrementally label samples in $\mathcal{D}_{u}$ to minimize the cost of annotation, better understand the annotation process, and ultimately improve model generalization with few labels. 

For each task, we initiate the AL process by acquiring $\mathcal{D}_{l}$ to train a shallow classifier $\mathcal{G}_{\omega}: \mathbb{R}^m \to \mathbb{R}^{|\mathcal{Y}|}$ on top of the $m$-dimensional representations from $\mathcal{F}_\theta$. $\mathcal{G}_{\omega}$ is a fully-connected linear model with softmax activation such that it outputs the probability an input $x \in \mathcal{X}$ belongs to each class $y \in \mathcal{Y}$:
\begin{equation}
    \mathcal{G}_{\omega}(\mathcal{F}_\theta(x)) = [P(y_1|x), P(y_2|x), \ldots, P(y_{|\mathcal{Y}|}|x)]. \label{eq:model_output}
\end{equation} 
Once $\mathcal{G}_{\omega}$ is trained on $\mathcal{D}_{l}$, model weights $\omega$ is fixed and predictions for each instance in $\mathcal{D}_{u}$ are generated. ALOE then uses an acquisition function $\mathcal{A}: \mathbb{R}^{|\mathcal{Y}|} \to \mathcal{X}$ to select the most valuable examples of $\mathcal{D}_{u}$ for label generation within a certain budget. Once these labels are acquired, this batch of examples is merged into the existing $\mathcal{D}_{l}$. It is followed by retraining $\mathcal{G}_{\omega}$ using the updated $\mathcal{D}_{l}$, and we repeat this process for a fixed number of AL acquisition steps. We note that instead of keeping $\mathcal{F}_{\theta}$ fixed during AL, one can fine-tune the entire model (i.e., $\mathcal{F}_{\theta}$ and $\mathcal{G}_{\omega}$) end-to-end, but this process can be very time-consuming as retraining has to be performed after each label acquisition step, and the learnable parameters also increase significantly. Similarly, the feature extraction model $\mathcal{F}_{\theta}$ will become task-specific; it may lose its generalizable nature and not perform well for other downstream tasks.

For the acquisition function $\mathcal{A}$, we use an uncertainty sampling-based method called \textit{smallest margin}, which is relatively simple yet shown to be effective in identifying informative examples for annotation \cite{settles2012active,lewis1994heterogeneous}. %
Specifically, $\mathcal{A}$ selects
\begin{align}
x^* = \arg\min_{x\in D_u}~P(y_{(1)}|x) - P(y_{(2)}|x), \label{eq:smallest_margin}
\end{align}
where $P(y_{(i)}|x)$ is the $i$th largest probability in \eqref{eq:model_output}, e.g. the predicted label for the sample is $y_{(1)}$. That is, \eqref{eq:smallest_margin} chooses samples that have a ``very close second prediction.'' Thus the model is less certain about them, and acquiring their labels from the oracle will provide more information about where the decision boundary of the updated model should be.

\setlength{\tabcolsep}{8pt}
\begin{table*}[t]
\centering
\footnotesize
\caption{\small{Comparison of \textit{test set} accuracy (\%) of our approach with other baselines on different end-tasks. ALOE achieves similar recognition rate with class-aware sampling, while using several folds less labeled examples. The baseline results are from  \cite{peplinski2020frill}, where available; else we train a linear classifier on time-averaged representations using published models.}}
\label{tab:main_table}
\begin{tabular}{@{}lcccccclc@{}}
\toprule
\multirow{3}{*}{\textbf{Dataset}} &
\multirow{3}{*}{\textbf{TRILL~\cite{shor20interspeech}}} &
\multirow{3}{*}{\textbf{TRILL-Dist~\cite{shor20interspeech}}} &
\multirow{3}{*}{\textbf{FRILL~\cite{peplinski2020frill}}} &
\multirow{3}{*}{\textbf{TRILLsson~\cite{shor2022trillsson}}} &
\multicolumn{4}{c}{\textbf{ALOE (Ours)}} \\ \cmidrule(l){6-9} 
&      &      &      &      & \multicolumn{2}{c}{\textit{Class-Aware}} & \multicolumn{2}{c}{\textit{Class-Agnostic}}                 \\
&      &      &      &      & \textit{Random}  & \textit{Uncertainty} & \multicolumn{1}{c}{\textit{Random}} & \textit{Uncertainty} \\ \midrule
MSWC (Micro-EN) & 81.3 & 74.4 & 79.1 & 93.7 & 90.6 & 93.0                 & 76.0  & 79.7  \\
SpeechCommands & 81.9 & 80.2 & 79.7 & 96.4 & 92.6 & 94.9 & 70.3 & 78.6 \\
Vocalsound  & 88.2 & 85.8 & 86.7 & 91.1 & 84.9  & 88.2 & 79.0  & 79.1   \\
Voxforge & 84.5 & 80.0 & 76.9 & 99.6 & 98.4 & 99.2 & 94.4 & 96.2          \\
FluentSpeech & 69.3 & 62.3 & 64.9 & 97.5 & 83.5 & 87.5 & 60.8 & 62.6 \\
\bottomrule
\end{tabular}
\end{table*}

\begin{figure*}[t]
\centering
\subfloat[Class-Aware Sampling]{
\includegraphics[width=0.18\textwidth]{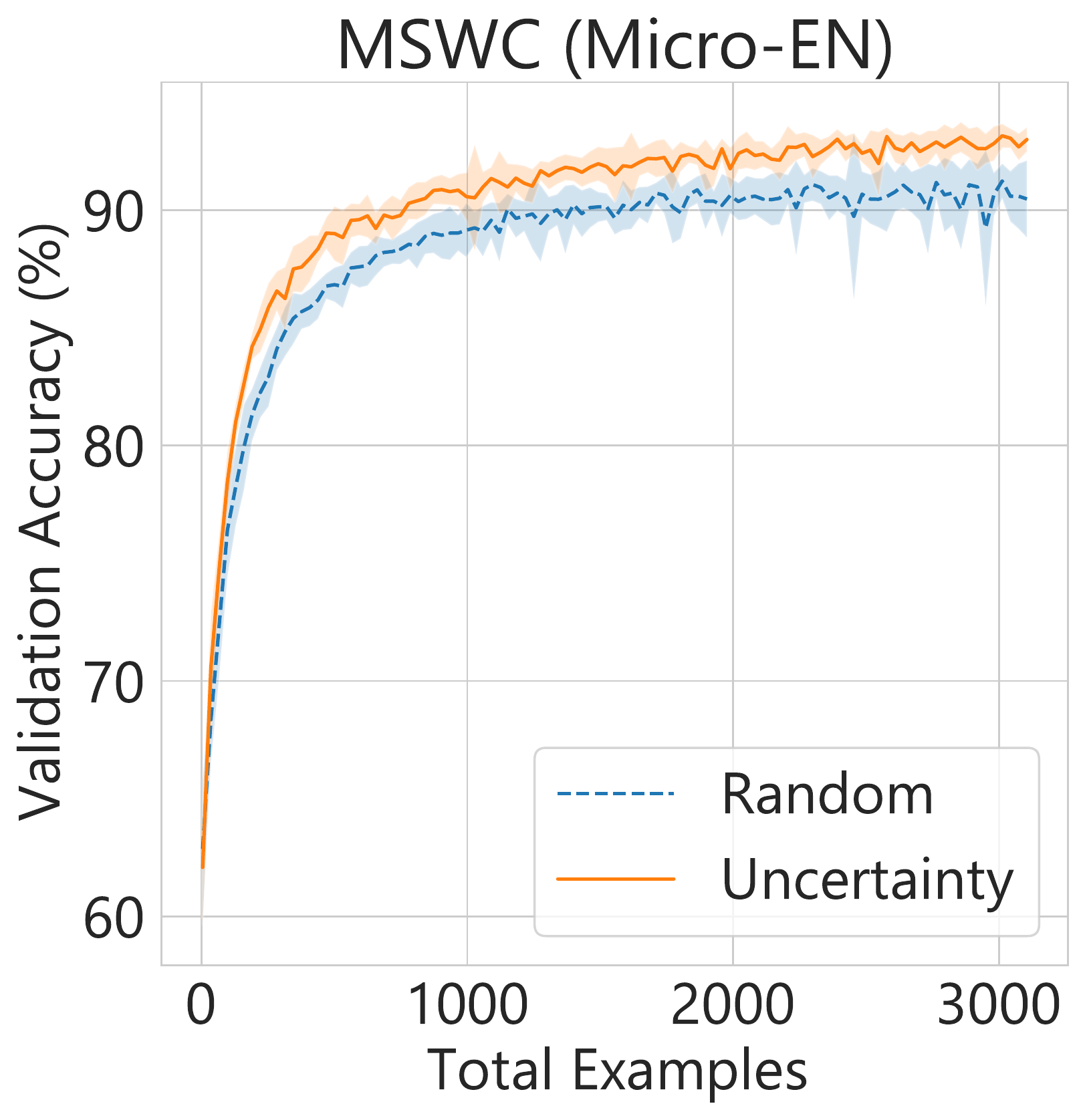}

\includegraphics[width=0.18\textwidth]{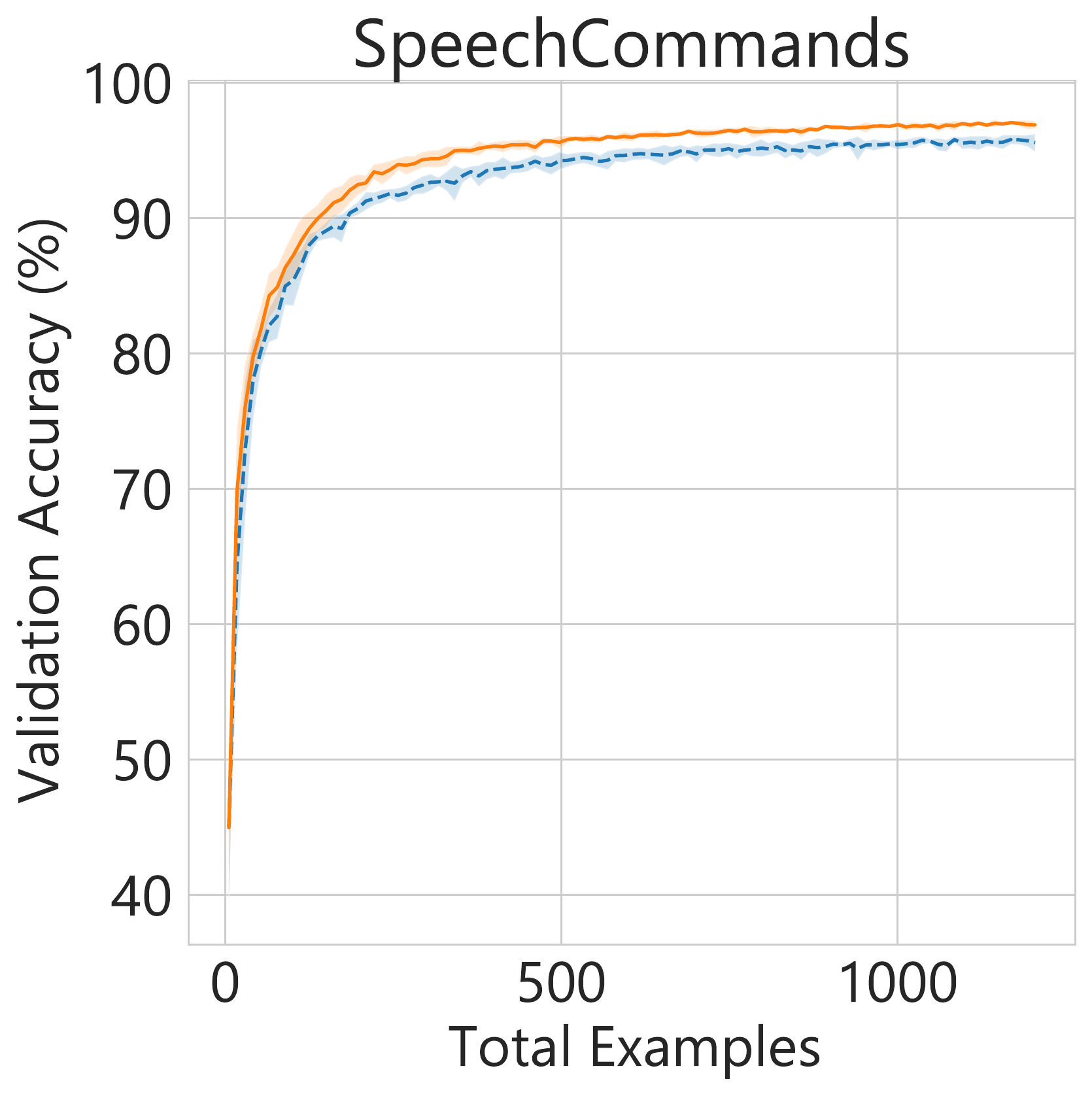}

\includegraphics[width=0.175\textwidth]{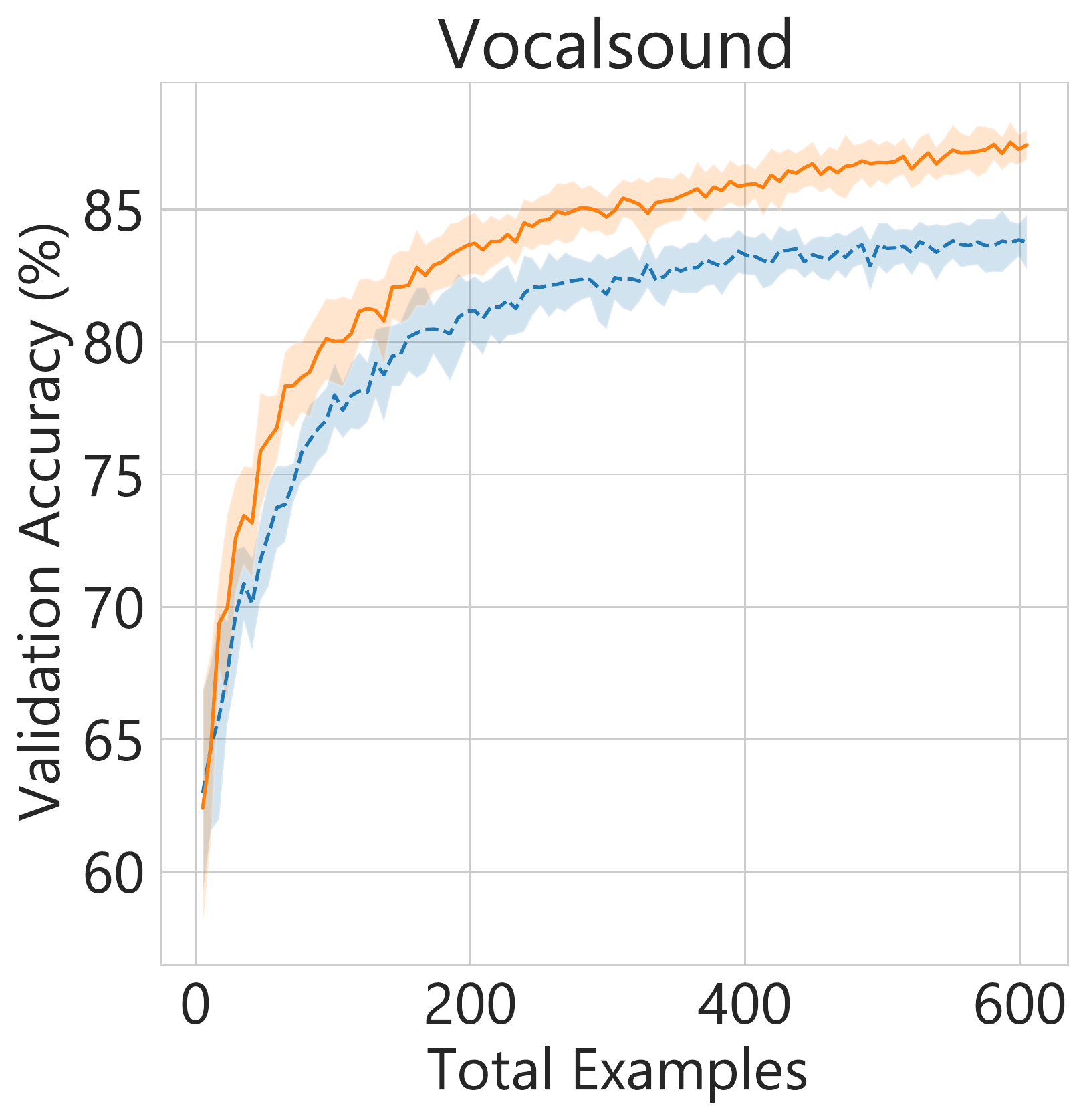}

\includegraphics[width=0.18\textwidth]{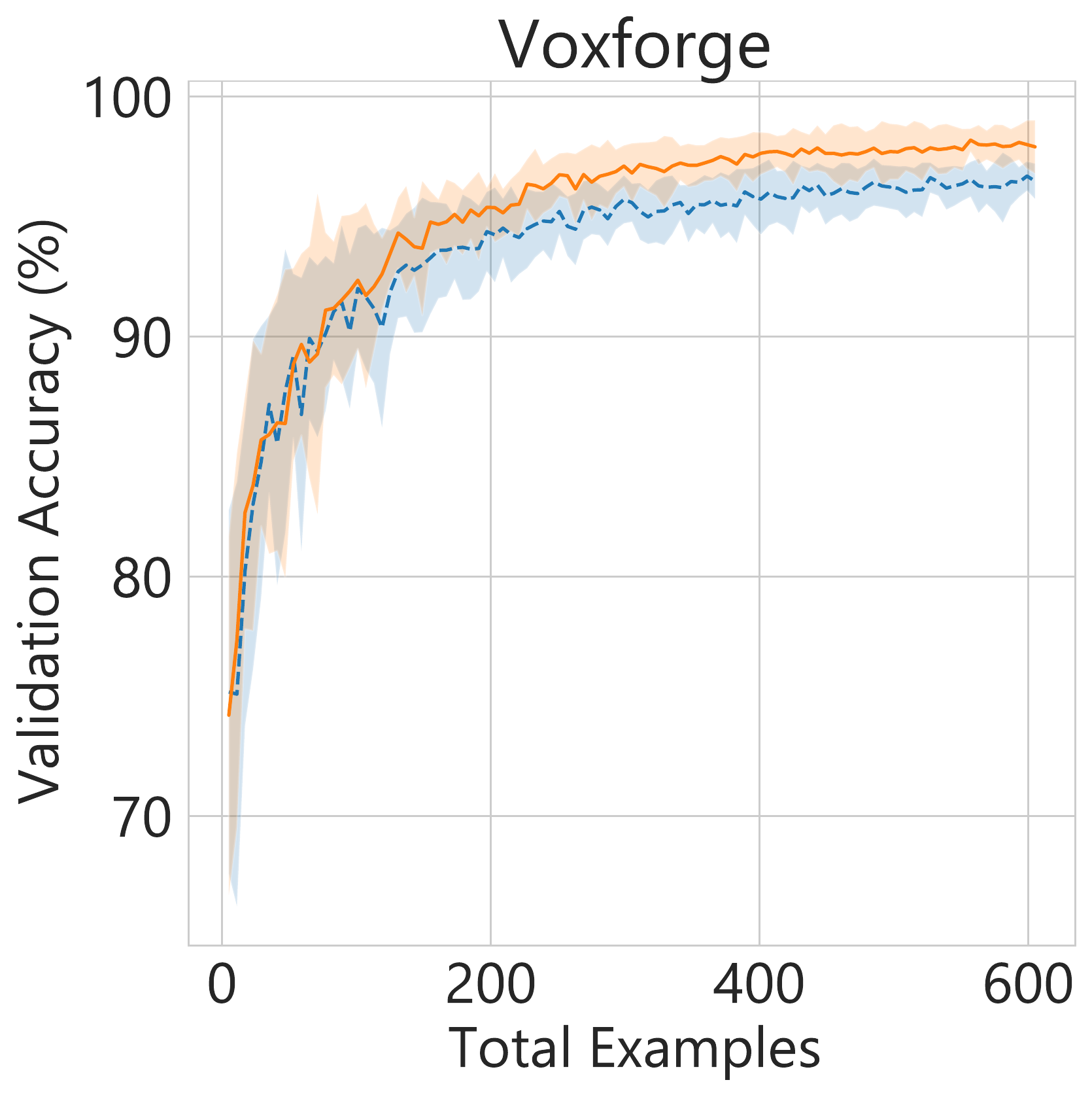}

\includegraphics[width=0.175\textwidth]{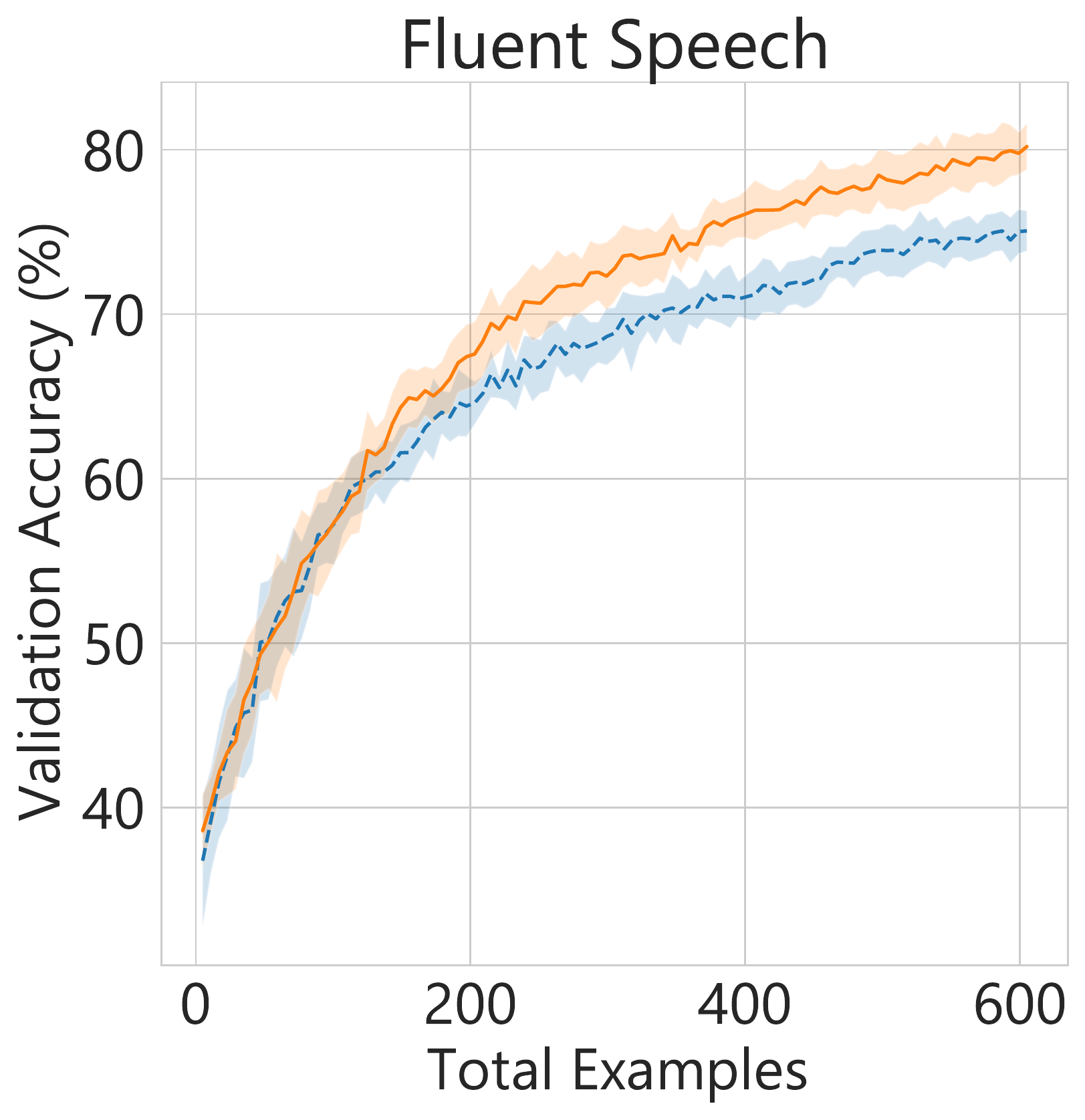}
}\\

\subfloat[Class-Agnostic Sampling]{
\includegraphics[width=0.18\textwidth]{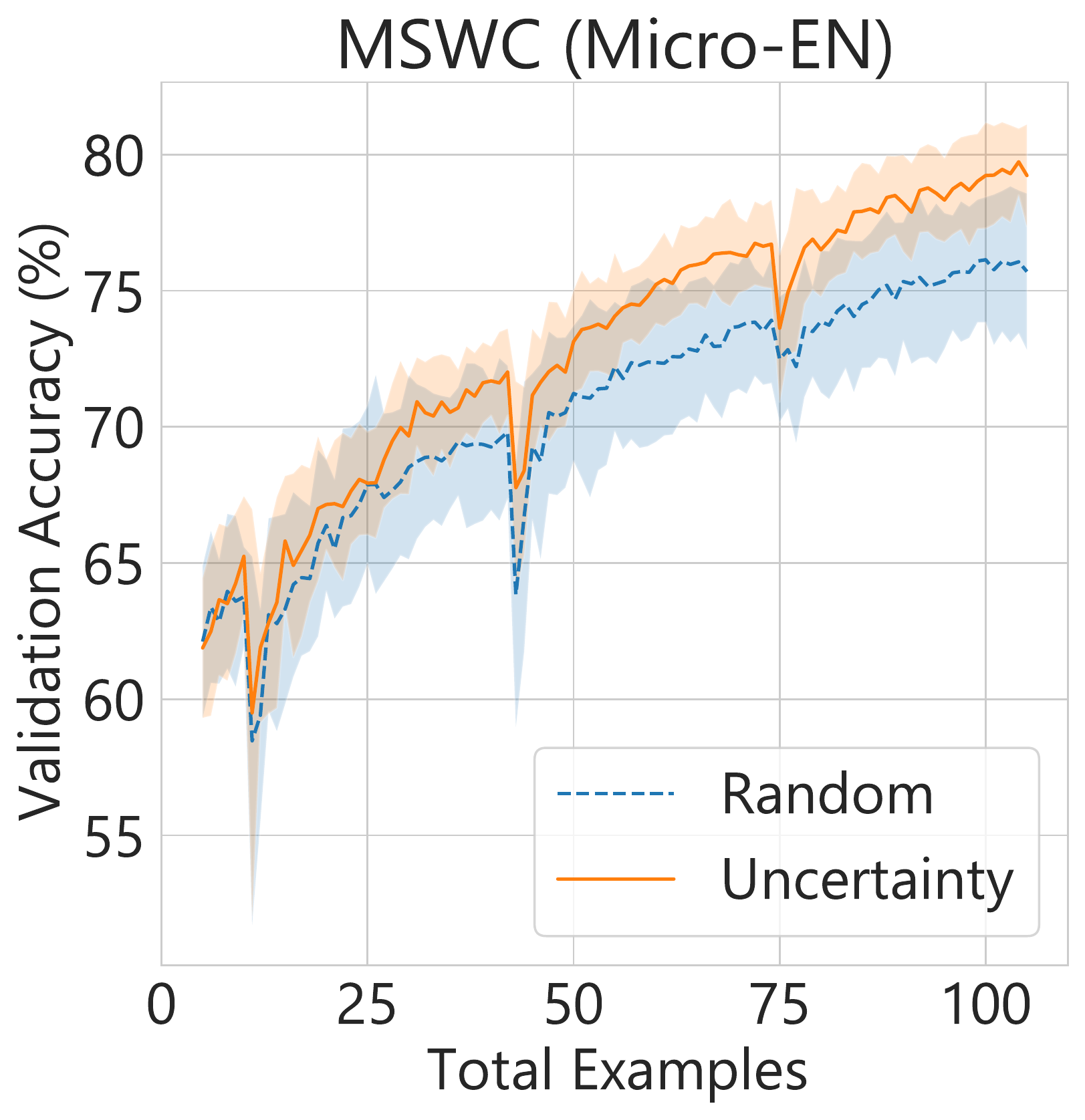}

\includegraphics[width=0.18\textwidth]{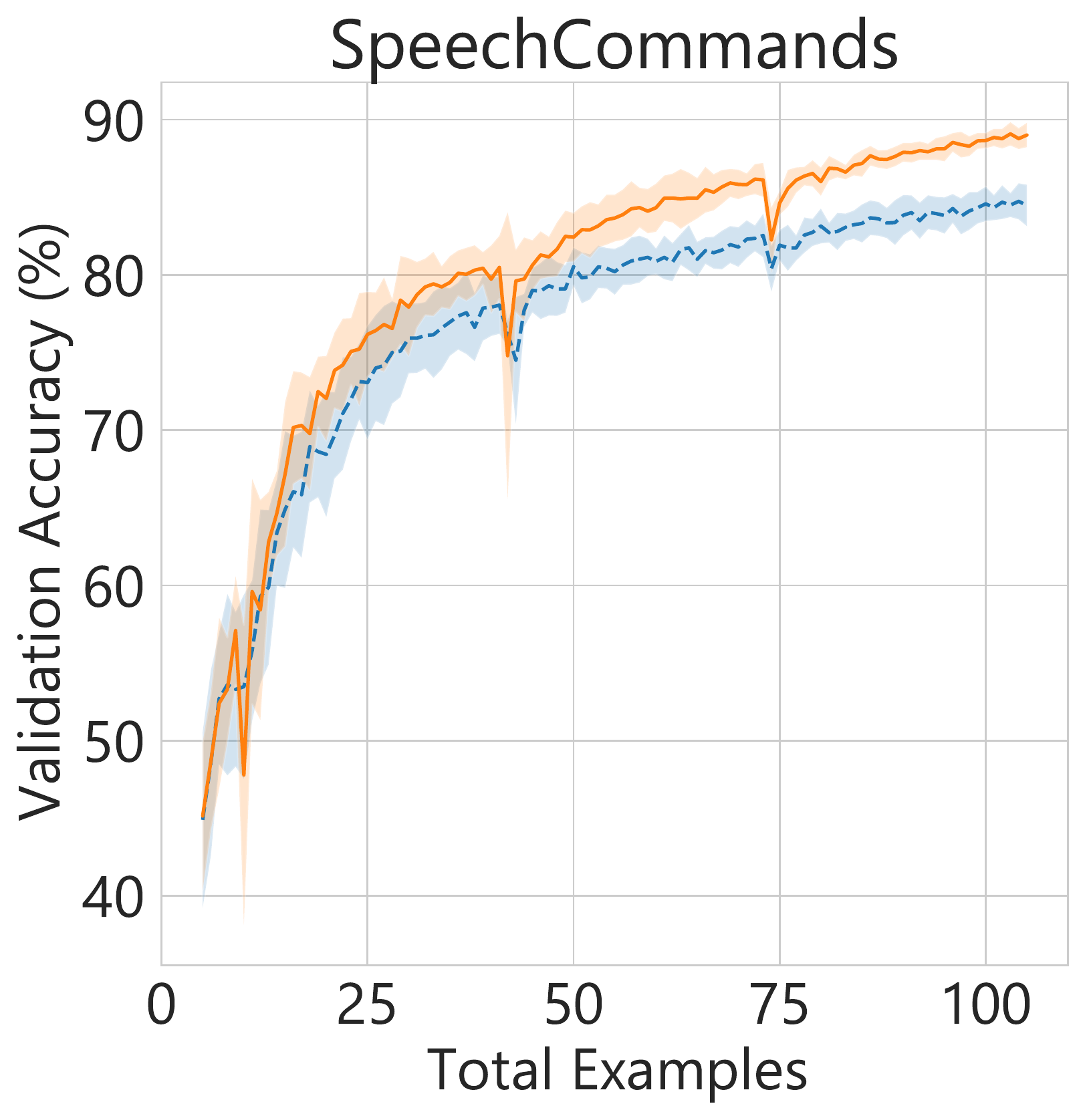}

\includegraphics[width=0.18\textwidth]{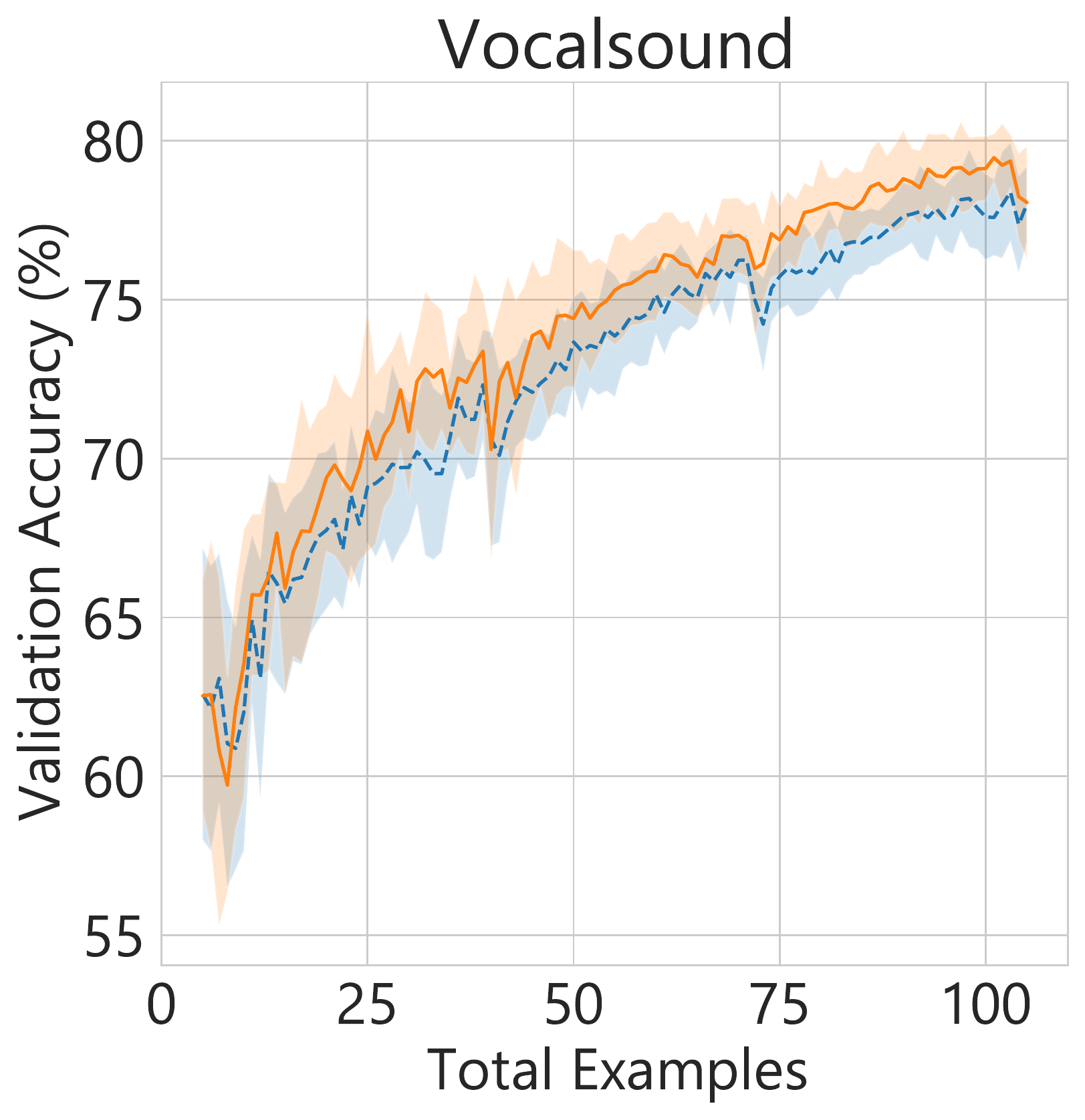}

\includegraphics[width=0.18\textwidth]{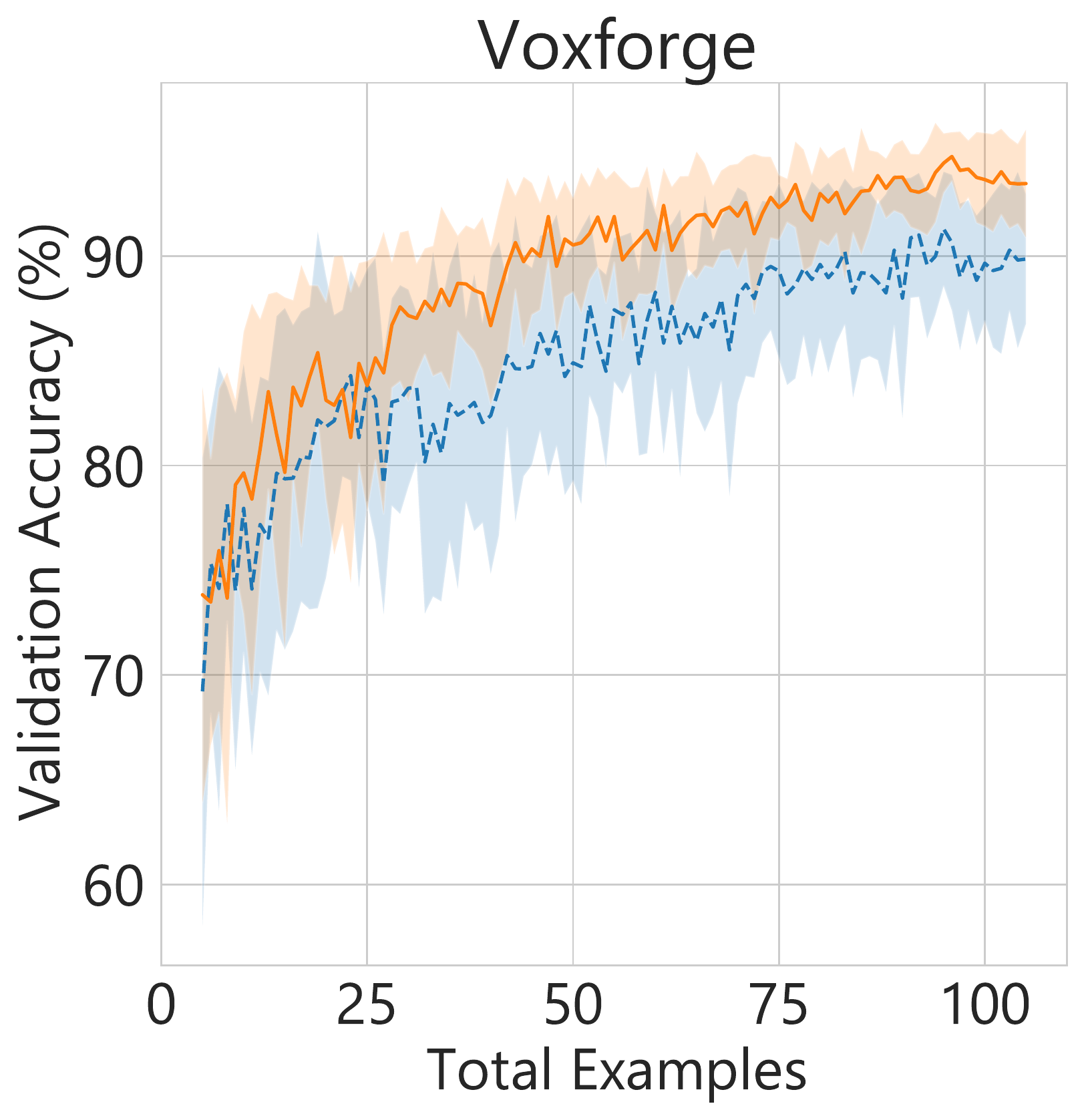}

\includegraphics[width=0.18\textwidth]{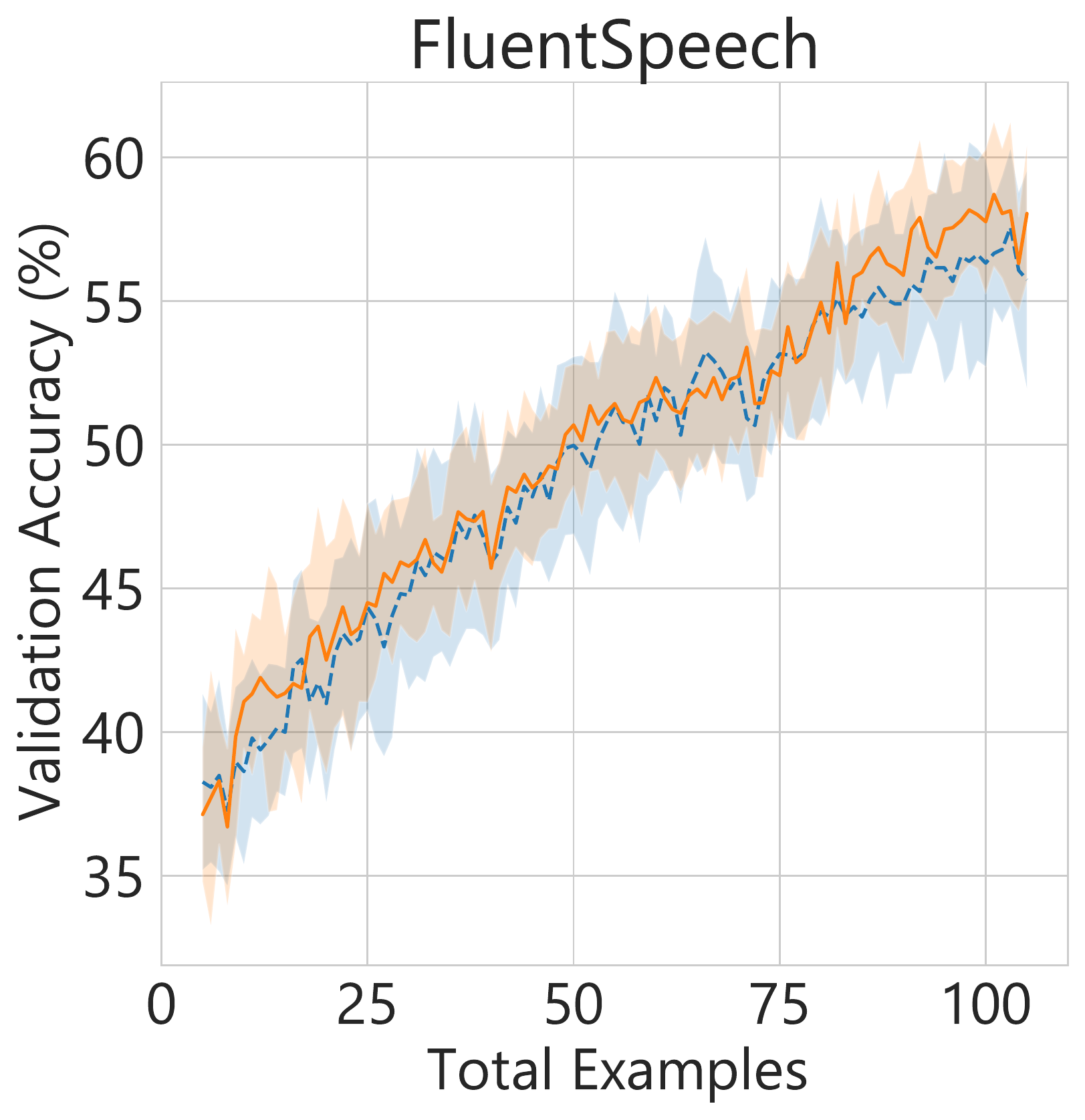}
} 
\caption{\small{Uncertainty sampling outperforms random sampling at every AL round, as measured by \textit{validation set} accuracy (\%). %
}}
\label{fig:al_budget}
\end{figure*}

\section{Experiments}
\label{sec:experiments}

For $\mathcal{F}_{\theta}$, we leverage publicly available pretrained models from the TRILLsson family, in particular, the Audio Spectrogram Transformer (AST) \cite{shor2022trillsson}. TRILLsson models are trained via knowledge distillation using large-scale unlabeled data with a massive self-supervised conformer-based teacher model. They provide state-of-the-art performance on a broad spectrum of downstream non-semantic speech tasks while being smaller than the teacher model. All pretrained models mentioned in this paper are from TensorFlow Hub~\cite{tensorflow2015-whitepaper}. $\mathcal{F}_{\theta}$ embeds the entire audio clip into a vector with dimension $m=1024$.

For each task, we report results aggregated from $10$ independent runs of the experiment. For each run of the experiment, there are $100$ AL acquisition steps, and at each AL acquisition step, $\mathcal{G}_{\omega}$ is trained for $100$ epochs with Adam and a learning rate of $0.001$. As the size of $\omega$ is quite small, the added benefit of our approach is that AL executes at a rapid pace. $\mathcal{D}_{l}$ is initially seeded with $5$ labeled examples per class. At each round of AL, \textit{class-aware} sampling selects one example per class based on the predicted labels, while \textit{class-agnostic} sampling picks a single example from all of $\mathcal{D}_u$. Note that this means they acquire $|\mathcal{Y}|$ labels and one label at each step, respectively. %

Several publicly available datasets are used to evaluate audio recognition models on non-semantic speech tasks~\cite{shor20interspeech, shor2022trillsson}. %
We perform experiments on MSWC (Micro-English) \cite{mazumder2021multilingual} and SpeechCommands \cite{warden2018speech} for {keyword spotting}, comprising of ($96,099$ samples, $31$ classes) and ($100,503$ samples, $12$ classes), respectively. For {spoken language identification}, we use Voxforge \cite{maclean2018voxforge} containing  $176,436$ audio clips from $6$ different languages. Similarly, for {human vocal sounds monitoring}, e.g., coughing and sneezing, we use Vocalsound \cite{gong_vocalsound} that has $21,024$ audio examples from $6$ classes. Finally, we evaluate on {action detection} ($30,043$ samples, $6$ classes) using FluentSpeech \cite{lugosch2019speech}. In all cases, we use the standard train, validation, and test splits provided with the original datasets, with audio sampled at $16$kHz.

We compare ALOE against many baseline models.
First, we investigate several other self-supervised or distillation-based methods. In TRILL \cite{shor20interspeech}, TRILL-Dist \cite{shor20interspeech}, FRILL \cite{peplinski2020frill}, and TRILLsson (AST), the linear classifiers are trained using the entire labeled data from the downstream task. These baselines, particularly the AST model, establish an upper-bound performance that can be achieved with a simple classifier trained with all labeled examples from a specific task. 
Additionally, we explore other uncertainty sampling methods \cite{settles2012active,miller2022graph} such as largest margin, least confidence, entropy, and norm,
as well as random acquisition, i.e. sampling from a uniform distribution on the pool $\mathcal{D}_{u}$ at each acquisition step.
Lastly, we compare the performance of AST as $\mathcal{F}_\theta$  to other neural network architectures such as ResNet-50 and EfficientNetv2 (B3) from the TRILLsson family \cite{shor2022trillsson}. Unless otherwise specified, ALOE uses the TRILLsson (AST) pretrained model with class-aware smallest margin sampling. 

\section{Results and Analysis}

\setlength{\tabcolsep}{14pt}
\begin{table*}[t]
\centering
\footnotesize
\caption{\small{Uncertainty sampling provides better performance on \textit{test set} than random sampling across different architectures.}}
\label{tab:arch_ablation}
\begin{tabular}{@{}lcccccc@{}}
\toprule
\multirow{2}{*}{\textbf{Dataset}} & \multicolumn{2}{c}{\textbf{ResNet-50}} & \multicolumn{2}{c}{\textbf{EfficientNet-v2 (B3)}} & \multicolumn{2}{c}{\textbf{Spectrogram Transformer (AST)}} \\ \cmidrule(l){2-7} 
& \textit{Random} & \textit{Uncertainty}    & \textit{Random} & \textit{Uncertainty}    & \textit{Random} & \textit{Uncertainty}    \\ \midrule
MSWC (Micro-EN) & 91.4\footnotesize{$\pm$ 0.70}  & \textbf{92.8\footnotesize{$\pm$ 0.58}} & 88.2\footnotesize{$\pm$ 1.11}   & \textbf{91.9\footnotesize{$\pm$ 0.56}} & 90.6\footnotesize{$\pm$ 1.48}   & \textbf{93.0\footnotesize{$\pm$ 0.50}} \\
SpeechCommands  & 92.4\footnotesize{$\pm$ 1.37}   & \textbf{93.1\footnotesize{$\pm$ 0.89}} & 90.4\footnotesize{$\pm$ 0.58}  & \textbf{92.5\footnotesize{$\pm$ 0.96}} & 92.6\footnotesize{$\pm$ 1.70}   & \textbf{94.9\footnotesize{$\pm$ 1.12}}  \\
Vocalsound      & 82.4\footnotesize{$\pm$ 0.92}  & \textbf{85.6\footnotesize{$\pm$ 0.80}} & 84.2\footnotesize{$\pm$ 0.73}  & \textbf{87.6\footnotesize{$\pm$ 0.38}} & 84.4\footnotesize{$\pm$ 0.94}  & \textbf{88.2\footnotesize{$\pm$ 0.35}} \\
Voxforge        & 95.7\footnotesize{$\pm$ 0.41}  & \textbf{97.4\footnotesize{$\pm$ 0.23}} & 97.0\footnotesize{$\pm$ 0.29}  & \textbf{98.4\footnotesize{$\pm$ 0.07}} & 98.4\footnotesize{$\pm$ 0.17}  & \textbf{99.2\footnotesize{$\pm$ 0.06}} \\
FluentSpeech & 79.0\footnotesize{$\pm$ 1.90}  & \textbf{82.7\footnotesize{$\pm$ 1.07}} & 80.9\footnotesize{$\pm$ 1.95}  & \textbf{85.1\footnotesize{$\pm$ 0.96}} & 83.5\footnotesize{$\pm$ 1.14}  & \textbf{87.5\footnotesize{$\pm$ 0.96}} \\

\bottomrule
\end{tabular}%
\end{table*}
\begin{figure}[t]
\centering
\subfloat{\includegraphics[width=0.44\columnwidth]{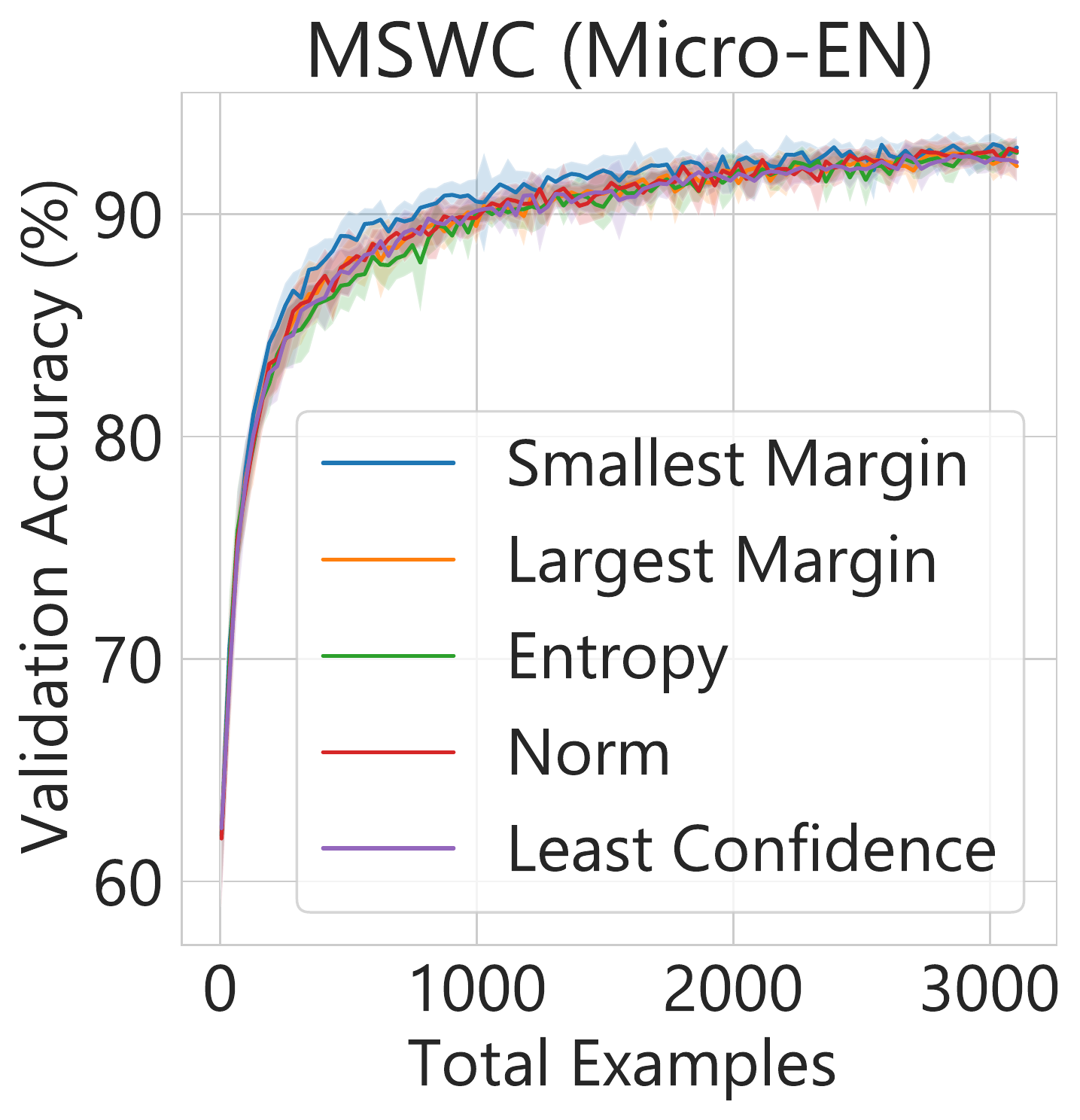}} 
\subfloat{\includegraphics[width=0.45\columnwidth]{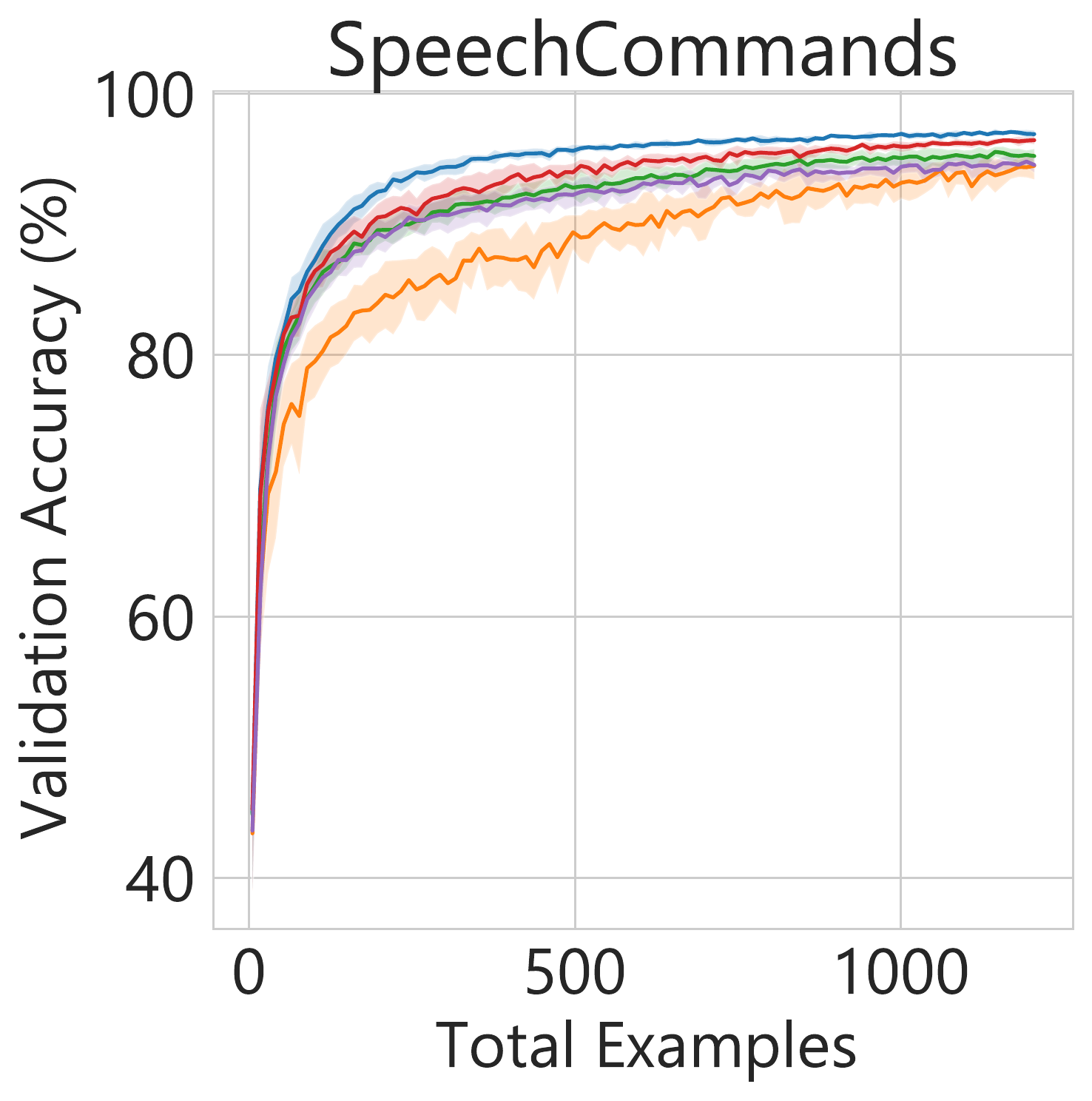}} \\
\subfloat{\includegraphics[width=0.44\columnwidth]{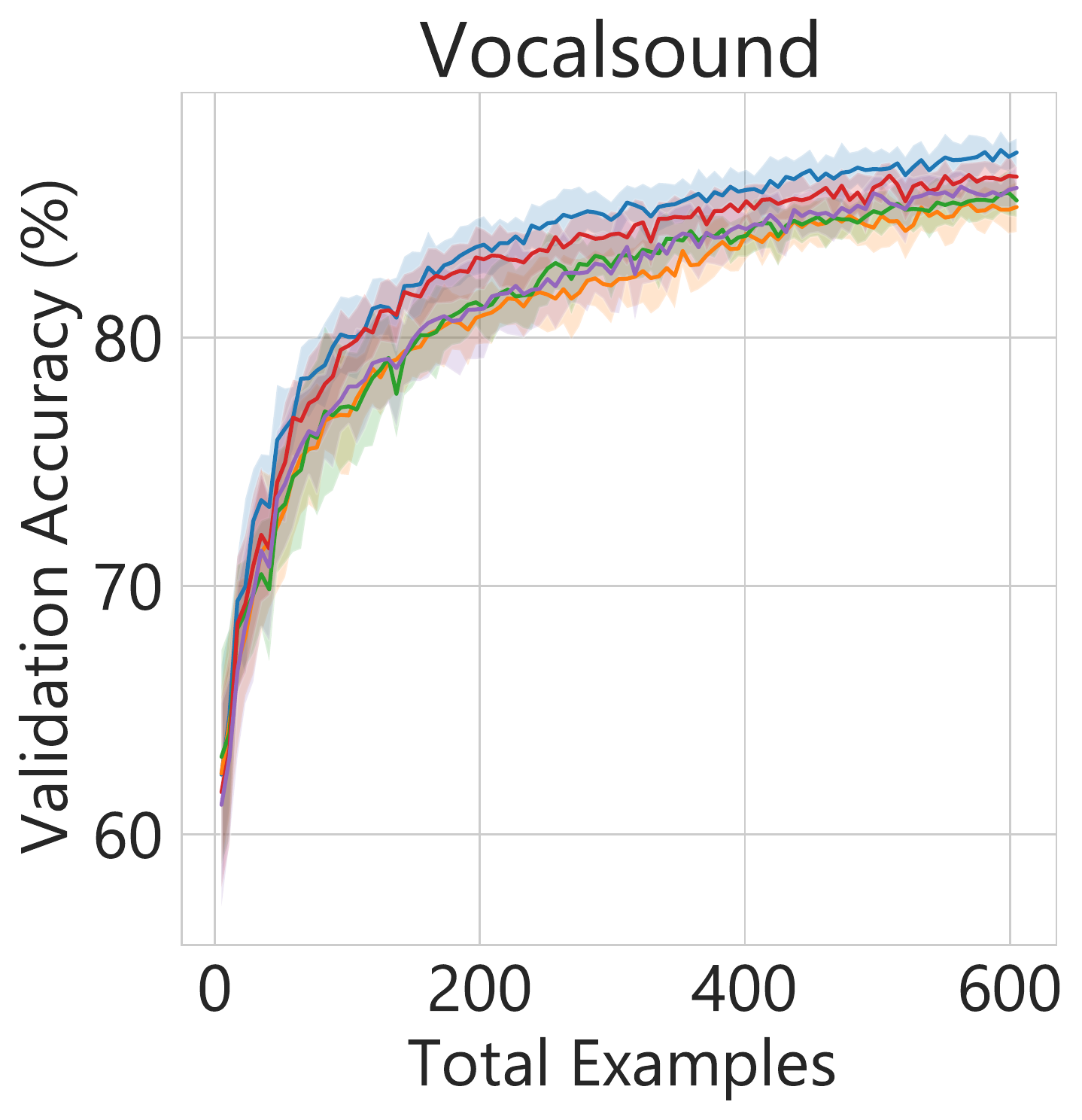}} 
\subfloat{\includegraphics[width=0.45\columnwidth]{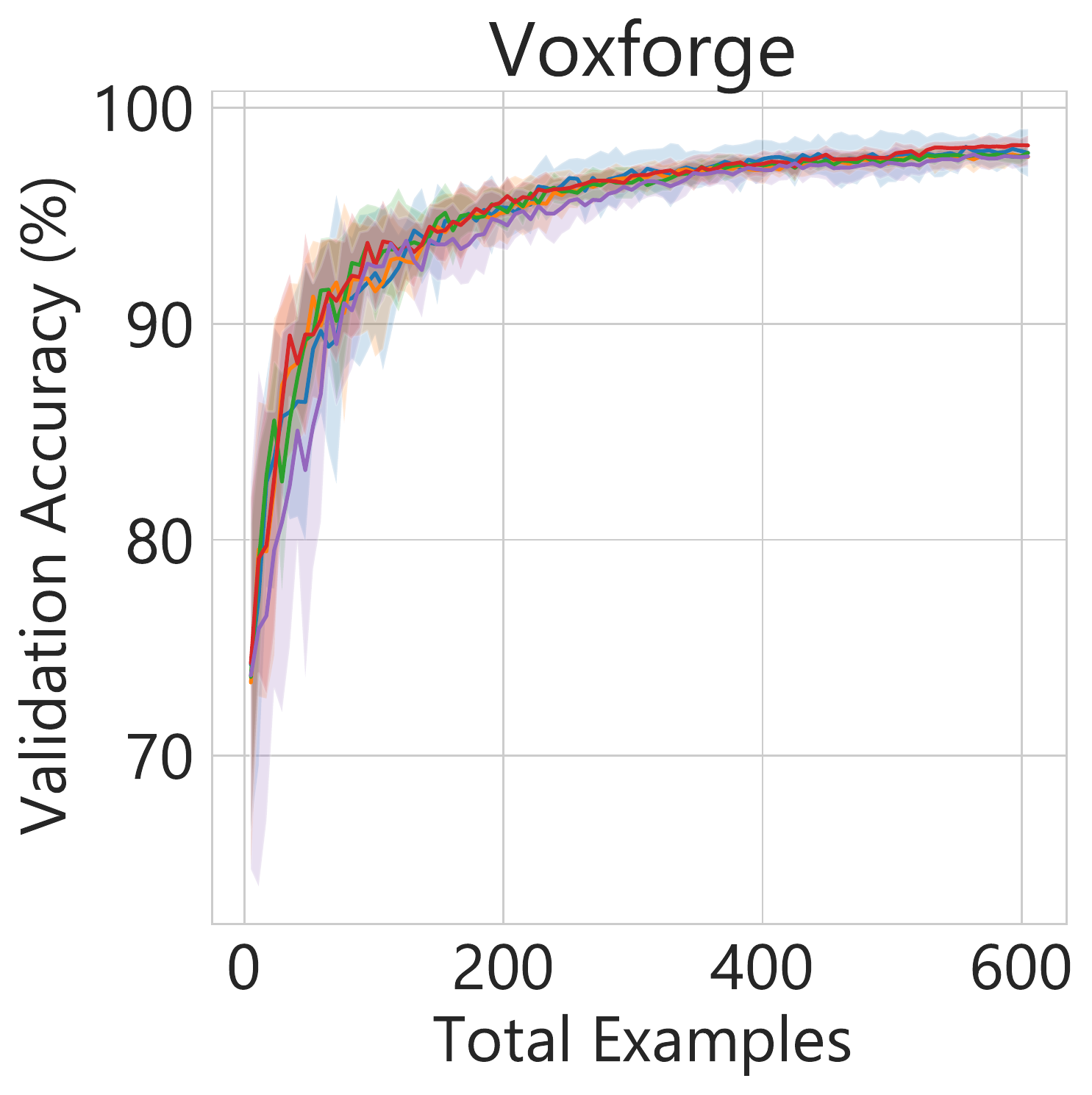}} 
\caption{\small{Different uncertainty sampling methods paired with pretrained model perform similarly well on \textit{validation set}.}}
\label{fig:sampling_methods}
\end{figure}
Table \ref{tab:main_table} compares the \textit{test set} performance of ALOE after 100 rounds of AL to other self-supervised or distillation-based models. As expected, TRILLsson has the highest recognition rate (often above 95\%) across a wide range of non-semantic speech classification tasks. But ALOE with class-aware uncertainty-based sampling achieves accuracy comparable to TRILLsson while using several times fewer labeled examples. For example, ALOE arrives at 99.2\% accuracy on Voxforge after only acquiring labels for 600 samples, surprisingly close to the 99.6\% achieved by the upper-bound model that used the entire labeled dataset. This suggests that with the pretrained TRILLsson (AST) model as $\mathcal{F}_\theta$, the linear classifier $\mathcal{G}_\omega$ needs only a couple hundred labeled examples to be near perfect in this task. The implications are impressive: when deployed in the real world, this human-in-the-loop approach may eliminate the cost of acquiring \textit{excess} labels in the first place, taking the guesswork out of data collection. We note that ALOE gets to within 0.7\% to 2.9\% of TRILLsson's accuracy in MSWC, SpeechCommands, and Vocalsound as well,  supporting the notion that AL with pretrained models can improve data- and label-efficiency in many non-semantic speech tasks.

Next, we discuss the effect of uncertainty-based sampling in ALOE. Comparing the neighboring random and uncertainty columns in Table \ref{tab:main_table} shows that although the extent may differ (0.1\% to 8.3\%), the uncertainty-based sampling method often leads to a higher test set accuracy for different tasks in both class-aware and class-agnostic settings. Figure \ref{fig:al_budget} confirms that this is also observed during the AL phase, as measured by validation set accuracy. These results indicate that ALOE successfully selects examples that are more informative than random picks. Within uncertainty-based sampling, the different acquisition functions perform similarly well; see Figure \ref{fig:sampling_methods}. Furthermore, we explore the effect of AST as the default pretrained model by switching it out with different neural network architectures. Results on the test set are summarized in Table \ref{tab:arch_ablation}. The different pretrained TRILLsson models performed similarly, but AST showed a slight advantage. We point out from Table \ref{tab:arch_ablation} that uncertainty sampling performs better than random sampling across different architectures as well.

Recall that class-aware setting selects more examples than class-agnostic setting at every acquisition step, and therefore these two cannot be directly compared to each other. Instead, users should choose an approach based on their AL labeling budgets and computational costs. For example, class-agnostic may make more sense when labeling is very expensive and there are many classes. Users also need to choose the number of AL acquisition steps. While we fixed that number at 100 for all experiments in this work, users can easily monitor validation accuracy as in Figure \ref{fig:al_budget} to determine stopping points. For instance, they may decide to do early stopping when the performance plateaus (e.g. SpeechCommands with class-aware sampling) or let the model train longer (e.g. Fluent Speech).

ALOE's core strength is in its simplicity. Indeed, adding more layers or complexity to $\mathcal{G}_\omega$ may improve the classification accuracy, but we wanted to show that even with a linear (i.e. single layer) model, optimal performance can be achieved. Also, it is standard protocol in self-supervised learning to evaluate pretrained models with linear probes \cite{saeed2021contrastive}. Further, keeping $\mathcal{G}_\omega$ small greatly improves training efficiency and reduces the number of parameters to be learned. 
To reiterate, ALOE does not fine-tune the encoder during AL as our motivation is to use a generic feature extractor for more than one task, and updating the parameters with few samples can have catastrophic consequences for the learned representations from large-scale data. However, we note that when a sizeable portion of the data is labeled at the end of the AL phase, it may be used for end-to-end model fine-tuning. 

\section{Conclusions} 
\label{sec:conclusions}
We proposed ALOE, a novel approach to improve label- and sample-efficiency of non-semantic speech tasks with active learning. It provides an end-to-end system that exploits pretrained models more effectively, consequently mitigating the need to prepare large labeled data for downstream tasks. ALOE's simplicity allows easy implementation, scaling and deployment. We plan to extend this framework to graph-based active learning and to AudioSet in the future.

\small
\bibliographystyle{IEEEbib}
\bibliography{main}

\end{document}